\documentstyle[prb,aps]{revtex}
\begin{document}

\newcommand{\tJm}{\mbox{$t$-$J$-model}}
\newcommand{\Gpipi}{\mbox{$\Gamma\to$($\pi$,$\,\pi$)}}
\newcommand{\Gpino}{\mbox{$\Gamma\to$($\pi$,$\,0$)}}
\newcommand{\pinonopi}{\mbox{($\pi$,$\,0$)$\to$($0$,$\,\pi$)}}
\newcommand{\pino}{\mbox{($\pi$,$\,0$)}}
\newcommand{\pipi}{\mbox{($\pi$,$\,\pi$)}}
\newcommand{\pizpiz}{\mbox{($\pi/2$,$\,\pi/2$)}}

\wideabs{
\title{Angle resolved photoemission spectroscopy of Sr$_2$CuO$_2$Cl$_2$ -- a revisit}

\author{C. D\"urr, S. Legner, R. Hayn, S. V. Borisenko, Z. Hu, A. Theresiak,
M. Knupfer, M. S. Golden, and J. Fink}
\address{Institute for Solid State and Materials Research Dresden, P.O.
Box 270016, D-01171 Dresden, Germany}
\author{F. Ronning and Z.-X. Shen}
\address{Department of Physics and Stanford Synchrotron Radiation
Laboratory, Stanford University, Stanford, CA 94305-4045, USA}
\author{H. Eisaki and S. Uchida}
\address{Department of Superconductivity, The University of Tokyo,
Bunkyo-ku, Tokyo 113, Japan}
\author{C. Janowitz and R. M\"uller}
\address{Institut f\"ur Physik der Humboldt-Universit\"at Berlin,
Invalidenstrasse 110, 10115 Berlin, Germany}
\author{R. L. Johnson}
\address{II. Institut f\"ur Experimentalphysik der Universit\"at Hamburg,
Luruper Chaussee 149, 22761 Hamburg, Germany}
\author{K. Rossnagel and L. Kipp}
\address{Institut f\"ur Experimentelle und Angewandte Physik,
Christian-Albrechts-Universit\"at zu Kiel,  Leibnizstrasse 19, 24118 Kiel, Germany}
\author{G. Reichardt}
\address{BESSY GmbH, Albert-Einstein-Str. 15, 12489 Berlin, Germany}

\maketitle

\begin{abstract}
We have investigated the lowest binding-energy electronic structure of the
model cuprate Sr$_2$CuO$_2$Cl$_2$ using angle resolved photoemission
spectroscopy (ARPES). Our data from about 80 cleavages
of Sr$_2$CuO$_2$Cl$_2$ single crystals give a comprehensive, self-consistent picture
of the nature of the first electron-removal state in this model undoped CuO$_2$-plane cuprate.
Firstly, we show a strong dependence on the polarization of the excitation light
which is understandable in the context of the matrix element governing the photoemission
process, which gives a state with the symmetry of a Zhang-Rice singlet.
Secondly, the strong, oscillatory dependence of the intensity of the Zhang-Rice singlet 
on the exciting photon-energy is shown to be consistent with interference effects
connected with the periodicity of the crystal structure in the crystallographic c-direction.
Thirdly, we measured the dispersion of the first electron-removal states along \Gpipi\ and
\Gpino, the latter being controversial in the literature, and have shown that the data are best 
fitted using an extended \tJm, and extract the relevant model parameters. 
An analysis of the spectral weight of the first ionization states for different excitation energies within
the approach used by Leung {\it et al.} (Phys. Rev. B {\bf 56}, 6320 (1997))
results in a strongly photon-energy dependent ratio between the coherent and
incoherent spectral weight.
The possible reasons for this observation and its physical implications are discussed.
\end{abstract}

}

\section{Introduction}

Extensive effort continues to be expended to understand the electronic
properties of the cuprate high $T_c$ superconducting compounds.  
Apart from many differences, the cuprate superconducting materials have
important properties in common. 
Namely they are all layered compounds possessing CuO$_2$-planes, where the origin
of the superconducting behavior is to be found. The electronic structure of
the states with the lowest binding energy is satisfactorily understandable
only in models which take into account correlation effects. 
ARPES has proven to be a
powerful tool to investigate the low binding-energy electronic structure of the cuprates.
Important examples for the success of this method are the mapping of the Fermi
surface \cite{Aebi94} and the determination of an anisotropy of the superconducting
gap consistent with a d-wave order parameter \cite{Shen95}.

In the undoped regime the cuprates are two-dimensional antiferromagnetic
insulators with CuO$_2$-planes resulting in a correlation-gapped half-filled
Cu3$d_{x^2-y^2}$-O2$p_x$/O2$p_y$
antibonding band of lowest binding-energy. Upon doping with holes, the states with
lowest binding energy evolve towards the chemical potential, and the cuprate
becomes metallic or superconducting. 
To clarify the main principles of the evolution of the lowest binding-energy states
with doping by applying model Hamiltonians, one has to go step by step from
simple situations or limiting cases to more complicated scenarios. A natural
starting point, then, is to investigate the propagation of a single hole in a
CuO$_2$-plane by studying the undoped parent compounds of cuprate superconductors
by photoemission, where one electron is removed from the CuO$_2$-plane.
The states thus investigated can be termed the first electron-removal states.

A special class of parent compounds are the oxyhalides such as Sr$_2$CuO$_2$Cl$_2$,
Ca$_2$CuO$_2$Cl$_2$, Sr$_2$CuO$_2$F$_2$ (one layer compounds), and
Ca$_3$Cu$_2$O$_4$Cl$_2$ (double layered compound) \cite{Miller90,Sowa90}, 
all of which contain an apical halogen atom, rather than an oxygen.
It is now well established that the apical oxygen which is contained in 
most of the cuprate superconductors is not necessary for high $T_c$
superconductivity as Hiroi {\it et al.} \onlinecite{Hiroi94} showed that
Ca$_{2-x}$Na$_x$CuO$_2$Cl$_2$ is superconducting ($T_c=26$K).
Out of the class of cuprate parent compounds without apical oxygen, the 
layer compound Sr$_2$CuO$_2$Cl$_2$ is remarkable due to its extremely stable
stoichiometry.
Despite considerable efforts so far it has not been possible to dope this substance
chemically. Furthermore, the CuO$_2$-planes are unbuckled and the crystal
structure shows neither orthorhombic distortion nor a superstructure
at least down to 10 K (Ref.  \onlinecite{Vaknin90}). 
Therefore, Sr$_2$CuO$_2$Cl$_2$ can be seen as the best realization of a
two-dimensional antiferromagnet at half-filling and it is in this limit an ideal
test case for any Hamiltonian in the low-doping regime. 

In theoretical studies, the lowest binding-energy ionization states of the cuprates were
predicted to be essentially described by a singlet antibonding combination of
a Cu3$d_{x^2-y^2}$ orbital (containing the intrinsic Cu3$d^9$ hole) and a coherent 
combination of the four neighboring O2$p_x$/O2$p_y$ orbitals which can be thought 
of as containing the hole created in the photoemission process \cite{Zhang88,Eskes90}.
This two-hole state is generally referred to as the Zhang-Rice singlet
state \cite{fer-zrs}.
The singlet character of the first electron-removal  states, at least in CuO, has been
experimentally verified using spin-resolved resonant photoemission\cite{Tjeng97}.

\par
There are various possibilities to theoretically model the dynamics of this lowest
binding-energy excitation. 
For instance, in the framework of a three-band Hubbard Hamiltonian ($H_{3b}$),
the Zhang-Rice singlet state is a two-particle state and
belongs to the $A_{1g}$ (totally symmetric) irreducible representation of the
eigenstates of $H_{3b}$.
In the one-band Hubbard Hamiltonian ($H_{1b}$), the \tJm\ and its extensions, the
Zhang-Rice singlet state is an
one-particle state belonging to the $B_{1g}$ irreducible representation and has the same
symmetry as a Cu3$d_{x^2-y^2}$ orbital.
Figure \ref{mirror} shows a schematic representation of a Zhang-Rice singlet state, including the two mirror 
planes M$_1$ and M$_2$ which are perpendicular to the CuO$_2$-plane and which are relevant 
for photoemission along along \Gpipi\ and \Gpino, respectively.

It is now generally accepted that the $H_{3b}$ is a good
starting point to describe the low binding-energy electronic structure in the
strongly correlated cuprate systems and consequently the $H_{3b}$ has been investigated in 
detail \cite{h3brefs}.
Additionally, it has been shown that the $H_{1b}$ still carries enough information to describe the 
low-energy dynamics in the cuprates \cite{h1brefs}.
In the next level of simplifying these models, the Hubbard Hamiltonian can be expressed
in terms of new Fermionic operators.
In the large-$U$-limit, this new Hamiltonian can be written in a $t^2/U$
series and gives an effective one-band Hamiltonian, the \tJm.
This has been shown to give the same results for both the
$H_{3b}$ (Ref. \onlinecite{Zhang88}) and the $H_{1b}$ (Ref. \onlinecite{Eskes94}) approaches.
Extensions of the \tJm\ have been carried out mainly in two different ways. 
Firstly, two further parameters - $t'$ describing the diagonal hopping and $t''$ for
the next nearest neighbor hopping - can be introduced in addition
to the direct hopping $t$.
\par
The fact that a $t$-mediated hopping creates strings of spin defects leads to a predicted
bandwidth for the Zhang-Rice singlet state which is governed by $t^2/U\sim J$.
Hopping events mediated by $t'$ and $t''$, however, take place on the same
spin sub-lattice and therefore even taking small values of these parameters 
has a significant effect on the results the \tJm\ gives for the $\mathbf{k}$-dependent
 spectral function of the Zhang-Rice singlet state.
Secondly, the \tJm\ has also been extended to take 
three-site hopping terms (proportional to $J/4$) into account  \cite{Eskes96},
which appear in a natural way if one employs a more detailed derivation of an
effective Hamiltonian for the low binding-energy excitations \cite{Yushankhai97}.

A different ansatz to understand the dynamics of holes in CuO$_2$-planes was
proposed by Laughlin \cite{Laughlin95}.
He argued that the hole created by photoemission decays into
spin and charge degrees of freedom and that the photoemission
experiment measures the dispersion of the spinon \cite{Laughlin95}.
However, in a gauge-field treatment of this model this decay of the quasiparticles is suppressed 
below the N\'eel-temperature, $T_N$, due to confinement \cite{Laughlin97}.
Another fundamental approach to calculate the properties of the
Hubbard Hamiltonian and the \tJm s uses the fact that they share an approximate SO(5)
symmetry \cite{MeixnerEder}.

The preceding discussion highlights the intense theoretical interest in the 
lowest binding-energy electron-removal states of the undoped cuprates.
Although originally trailing a few years behind the theoretical work, experimental
investigations of the first electron-removal  state in undoped cuprates have been carried
out using ARPES \cite{Wells95,LaRosa97,Kim98,Ronning98,Haffner00}.
Before going on to mention the status in the field to date, we first ask the question:
what kind of information does an ARPES experiment provide us with?
\par
Neglecting ``extrinsic'' effects such as scattering of the photoelectrons on their way to the surface
and assuming the applicability of the sudden approximation, the photoemission intensity $I({\mathbf k},E)$, 
reads

\begin{equation}\label{photointensity}
I({\mathbf k},E)\sim\sum_{\mid i\rangle,\mid f\rangle}
\mid M_{i,f}\mid^2\,S({\mathbf k},E)\,f(E)
\end{equation}

where the sum runs over all final states and initial states,
$f(E)$ denotes the Fermi function and $S({\mathbf k},E)$ the spectral function:

\begin{equation}\label{spectralweight}
S({\mathbf k}, E)=-\frac1\pi\,\mbox{Im}\,\langle i\mid c_{\mathbf
k}\,\frac1{E-H+E_i}\,c^\dagger_{\mathbf k}
\mid i\rangle\quad .
\end{equation}

$M_{i,f}=\langle f\mid{\mathbf A}\cdot{\mathbf p}\mid i\rangle$
is the matrix element to be taken between the initial state $\mid
i\rangle$ with energy $E_i$ and final states $\mid f\rangle$ with energy $E$,
${\mathbf A}\cdot{\mathbf p}$ is the photoemission interaction operator and
$H$ the Hamiltonian of the system.
From an ARPES experiment one can gain mainly two kinds of information. 
Firstly, the spectral function $S({\mathbf k},E)$ gives direct information
about the dispersion and the quasiparticle character of the states and can be
compared directly to predictions of model Hamiltonians. All information carried by
$S({\mathbf k}, E)$ is expected to be {\it independent} of the excitation
energy and geometry of the experiment.
Secondly, the matrix element $M_{i,f}$ includes all information concerning the
photoemission interaction. 
As such it is sensitive to the experimental geometry via symmetry
selection rules as well as to the photon-energy.

\par
The first ARPES experiments on Sr$_2$CuO$_2$Cl$_2$ (Ref. \onlinecite{Wells95}) were able
to make the important 
observation that the Zhang-Rice singlet state bandwidth along \Gpipi\ is of the order of 2$J$, and that 
the Zhang-Rice singlet state approaches closest to the chemical potential at the \pizpiz\ point.
These observations fit the predictions of the \tJm.
However, these early experiments found very little dispersion along \Gpino,
which is in conflict with the predictions of the same model.
Subsequent studies have confirmed the behavior along \Gpipi,
but have also observed dispersion along the \Gpino -direction in 
reciprocal space ($\mathbf{k}$-space)\cite{LaRosa97,Kim98,Haffner00}.

\par
Nevertheless, the situation as regards the exact dispersion relation along \Gpino,
and in particular concerning the 
minimum energy difference between the states along \Gpino\ and those at the \pizpiz\
point is unclear. This is more than a mere ARPES detail, as it is along
the \Gpino\ direction that the predictions of the different
models vary the most - thus making this direction in $\mathbf{k}$-space important for the
quantitative comparison between theory and experiment.

\par
Furthermore, the recent controversy surrounding the correct Fermi surface topology
in the doped high temperature superconductor
Bi$_2$Sr$_2$CaCu$_2$O$_{8+\delta}$ (Ref. \onlinecite{Chuangandsoon})
has illustrated that data sets recorded
from the same system with different experimental conditions can be remarkably dissimilar.
Consequently, both the photon-energy and exact polarization geometry used in an ARPES
experiment are 
important parameters which, if they cannot be treated at a quantitative, microscopic level,
should at least be thoroughly investigated on the experimental side.

\par
In this paper, we address the electronic structure and dynamics of the lowest binding-energy
electron-removal states
in the ``standard'' undoped model cuprate Sr$_2$CuO$_2$Cl$_2$ using ARPES.
Following the description of the experimental status given above, our ``re-visit''
of this system concentrates on the following points:

a) a thorough characterization of the photon-energy dependence of the first electron-removal
states at \pizpiz\ and close to \mbox{($\pi/2$,$\,0$)}
(the two points along \Gpipi\ and \Gpino\ for which the first electron-removal states
have maximum spectral weight);
b) a thorough characterization of the polarization dependence of the first electron-removal 
states along these high-symmetry directions in $\mathbf{k}$-space;
c) the determination of the dispersion relation of the first electron-removal  states along 
\Gpipi\ and \Gpino, with subsequent comparison of the results with both existing ARPES data 
and the predictions of an extended \tJm\ and 
d) the determination of the $\mathbf{k}$-dependent evolution of the coherent and incoherent parts of the
spectral weight of the first electron-removal  state along \Gpipi\ and \Gpino, and comparison 
of these data with theoretical predictions.

\section{Experiment}

Experiments were performed at the beamlines F$2.2$ and W$3.2$ at the
Hamburg Synchrotron Laboratory (HASYLAB), at the undulator beamline 
U$2$-FSGM and the $2$m SEYA beamline at the Berliner Elektronenspeicherring
Gesellschaft f\"ur Synchrotronstrahlung mbH (BESSY). At the storage rings and
monochromators used in these studies, highly linearly polarized synchrotron radiation
was available. In addition, the crossed undulator U$2$-FSGM beamline gave
the possibility to use vertically oriented linearly polarized light.
At this latter facility, polarization-dependent measurements could be performed without changing 
any other parameter in the experiment.
A total energy resolution (beamline and analyzer) of better than $70$ meV and an angular
acceptance of the analyzer of $\pm1$ degree was used. Fresh samples of
Sr$_2$CuO$_2$Cl$_2$ were cleaved {\it in situ} at a base pressure of
$1\times10^{-10}$mbar, and spectra were taken within six hours after
cleavage. Samples were either pre-oriented using X-ray diffraction measurements
or aligned {\it in-situ} with the aid of low energy electron diffraction (LEED). In all
cases, the fine
angular adjustment was carried out using the $\mathbf{k}$-space symmetry of the sharp ARPES features 
related to non-bonding O2$p$ states around \pipi\ and \pino, as discussed in
Refs. \onlinecite{Pothuizen97} and \onlinecite{Hayn99}. 

\par
The high-quality single crystalline Sr$_2$CuO$_2$Cl$_2$ was grown according
to the method described in Ref.  \onlinecite{Miller90}, where pre-dried high-purity
SrCO$_3$, SrCl$_2$ and CuO in a ratio 1:1:1 were melted at 1100$\,^\circ$C.
Although T$_N$ for Sr$_2$CuO$_2$Cl$_2$ is $251$ K (Ref. \onlinecite{Vaknin90}), 
all ARPES experiments were carried out at room temperature.
This decision was based on two arguments.
Firstly, and most importantly, as these experiments involve electron
ejection from perfect single crystals of a compound with an energy gap 
of the order of 2 eV, we had to eliminate uncertainties due to 
charging effects, which meant measuring at room temperature.
Secondly, although we measure at a temperature some 50 K above the 
N\'eel temperature, it has been shown that the antiferromagnetic 
spin correlation length of Sr$_2$CuO$_2$Cl$_2$ at $350$ K is still $250$
\AA\ (Ref. \onlinecite{Greven94}),
meaning that the lowest binding-energy hole state created in the photoemission process is 
still embedded in an antiferromagnetic spin background.

To be absolutely sure that variations in the flux of the incident photons from the
storage ring did not
lead to charging-induced energy shifts, we also adjusted the beamline such that
the photon flux impinging on the sample (monitored with a gold mesh upstream of the sample)
was constant.
This means that measuring time is the only parameter required for the normalization
of the data. Bearing in mind that the data presented here are a selection of
data from about 80 different cleavages, we made the qualitative observation
that best and sharpest energy distribution curves (EDC) were observed for samples
which behaved most sensitively to charging effects. This relation, of course, is
reasonable as it correlates cleavage quality with the intrinsically highly insulating nature 
of perfect Sr$_2$CuO$_2$Cl$_2$.

Essentially two different experimental geometries were used. They can be
characterized by the orientation of two planes: a) the plane of polarization
and b) the emission plane. The plane of polarization is the plane spanned by
the vector of the direction of the synchrotron light and the vector of its
electric field (vector of polarization). The emission plane includes the
vector of the direction of the photoelectron and the surface normal.
Throughout this paper we call the experimental geometry {\it parallel} if
these two planes are parallel, {\it perpendicular} if these two planes are
perpendicular to each other (see Fig. \ref{geometry} for a sketch of the two geometries). 

\section{Photon-energy and polarization dependence of the first electron-removal
states of an undoped CuO$_2$-plane}

As mentioned above, both the inconsistencies in the literature regarding the dispersion
relation of the first electron-removal  states along the \Gpino\ direction 
(Ref. \onlinecite{brillouin})
in Sr$_2$CuO$_2$Cl$_2$, as
well as the ongoing controversy regarding the photon-energy dependence of the ARPES
data from Bi$_2$Sr$_2$CaCu$_2$O$_{8+\delta}$, mean that before comparison with theory is
carried out, the 
photon-energy dependence of the first electron-removal  states should be examined in detail.

\par
In Fig. \ref{hnscan1}, photon-energy dependent ARPES data for the first electron-removal 
state in Sr$_2$CuO$_2$Cl$_2$ at two points in the Brillouin zone: a) \pizpiz\
(perpendicular geometry); and b) \mbox{($0.7\pi$,$\,0$)} (parallel geometry) are shown.
In these regions of $\mathbf{k}$-space the spectral weight of the first electron-removal  state is known
to be at a local maximum along the respective $\mathbf{k}$-space directions \cite{Wells95,LaRosa97,Kim98,Haffner00}.  
The spectra, which are normalized as described in the last section,
illustrate clearly the strong variation of the first electron-removal  state intensity with 
photon-energy. 
The data shown in Fig. \ref{hnscan1} represent only a small portion of the photon-energy dependent data 
recorded, and are intended to give the reader a direct impression of the strength of the effects at play. 
Each of the shown spectra is part of a short
k$_\parallel$-series of three or five spectra.
This was done to ensure we always captured the spectrum with the highest
spectral intensity for the first electron-removal states. Within the errors given by the finite $\mathbf{k}$-resolution 
of the experimental
setup ($0.054$ \AA$^{-1}$ for 16 eV photon-energy up to $0.152$ \AA$^{-1}$ for 80 eV
photon-energy)
we found the highest intensities always at the same $\mathbf{k}$-positions,
namely at \pizpiz\ and at \mbox{($0.7\pi$,$\,0$)}.

This is in contrast to the results reported in Ref. \onlinecite{Haffner00}, where
series of EDCs on Sr$_2$CuO$_2$Cl$_2$ along \Gpipi\ taken with two different
photon-energies show differences not only in intensity, but also in the
EDC-derived dispersion relation of the first ionization states.
This was discussed in Ref. \onlinecite{Haffner00} as {\it not} being due to either a) the experimental setup;
b) the excitation of different initial states or c) in the variation
of k$_\perp$ of the photoelectron. 
The effect was rather attributed to the strong impact of the matrix element, not only on the strength of the photoemission
signal but also on the position of the quasiparticle
peak in the photoemission spectra, leading to differences in the EDC-derived
dispersion relations.
We are forced to disagree with the last point, as in our extensive collection of ARPES data, there was never 
evidence for a photon-energy dependent shift of the $\mathbf{k}$-position for which the first
electron-removal states have minimum binding energy along \Gpipi.
The same, in fact, holds for other directions in the Brillouin zone as well as for the dispersion of other features with low binding energy. 
If matrix element effects change the EDC-derived dispersion relation,
then this change is, at least for our data, smaller than
the energy resolution and angular resolving power of our ARPES experiments, which is equal to
$\Delta_k$ or better than $\Delta_E$ the values used in Ref. \onlinecite{Haffner00}).

\par
Figure \ref{hnscan2} shows an analysis of the spectral weight of the first electron-removal
states at the same $\mathbf{k}$-points as shown in Fig. \ref{hnscan1}.
These data are derived from a large set of photon-energy dependent ARPES data
covering measurements from 10 (6) cleaves for the upper (lower) panels.
From Fig. \ref{hnscan2} it is clear that the ARPES spectral weight of the first
electron-removal  states oscillates with the final state kinetic energy, that is with k$_\perp$. 
We used a value of $8.0$ eV for the inner potential $E_0$
($E_0=V_0-\Phi$, where $\Phi$ is the work function) to calculate k$_\perp$,
a value which in the range of $6.9-8.9$ eV used by other
groups \cite{Courths84}. We see clear maxima at k$_\perp$=$0.82$, $1.63$, $2.40$
and $3.12$ \AA$^{-1}$, corresponding to photon-energies of $16$, $25$, $35$ and $48$ eV.

The oscillatory nature of the photon-energy dependence, coupled to the absence of a classical
resonance behavior at the Cu 3p threshold (around 74-76 eV photon-energy) indicates
that the factor dominating the observed behavior is something other than the atomic
photoionization cross-sections, and could be related,
for example, to the extreme two-dimensionality of the electronic states concerned.
This could lead to a matching of the final state k$_\perp$ to the
periodicity of the unit cell of Sr$_2$CuO$_2$Cl$_2$ in c-direction. Note that
the differences between the four k$_\perp$ is $0.72-0.81$ \AA$^{-1}$ which represents
in real space a distance of $7.8-8.7$ \AA\ which is comparable to c-axis separation of
two neighboring CuO$_2$-planes ($=7.805$ \AA) in Sr$_2$CuO$_2$Cl$_2$. The oscillation
of the photoemission intensity with photon-energy can therefore be attributed to interference effects of the
photoelectron wave diffracted from the c-axis periodicity of the layered crystal structure, similar to the
explanation of the strong photon-energy dependence of the photoemission intensity from the molecular orbitals of C$_{60}$ 
(Ref. \onlinecite{BerlinguysC60}).

These strong variations in intensity as a function of photon-energy are also reminiscent
of ARPES
data of YBa$_2$Cu$_3$O$_{7-\delta}$, in which an out-of-phase behavior regarding intensity
{\it vs.} k$_\perp$ was 
observed for the CuO$_2$-derived band which crosses the Fermi energy
near the $\bar{\mbox{X}}$ point and the so-called ``1 eV peak'' \cite{Tobin92}.
This fact was, at that time, used to argue against the surface-state origin of the 1 eV
peak, a feature which is now believed to be due to non-mixing O2$p$ states of particular
symmetry  \cite{Pothuizen97}. 
Nevertheless, the data presented here, taken together with the theoretical and experimental
investigations on Bi$_2$Sr$_2$CaCu$_2$O$_{8+\delta}$ (Refs. \onlinecite{Bansil99} and \onlinecite{Chuangandsoon}),
indicate clearly that care should be taken in the interpretation of absolute spectral weights observed in
the ARPES data of the layered cuprates, as matrix 
element and diffraction effects do play an important role in these systems.

A further experimental variable in an ARPES experiment is the polarization of the incoming
radiation. Figure \ref{series} shows two series of ARPES measurements on
Sr$_2$CuO$_2$Cl$_2$ recorded along the high-symmetry directions \Gpino\ and \Gpipi. 
The photon-energy was set to $22.4$ eV, which is near a maximum of intensity as shown in Fig. \ref{hnscan1} above. 
In each case, the series are presented in pairs of data sets recorded with perpendicular and parallel polarization 
geometries as described in the experimental section.

Along the \Gpipi\ direction in $\mathbf{k}$-space the first electron-removal state feature shows
highest photoemission intensity with perpendicular geometry (Fig.
\ref{series}a), whereas along \Gpino\ maximal intensities are observed in the 
parallel geometry (Fig. \ref{series}d), a result which tallies with earlier measurements with 25
eV photon-energy \cite{LaRosa97}.
This remarkable dependence of the photoemission intensity on
the polarization was first explained in the context of the strong polarization dependence
in photoemission data from surface states  \cite{Gobeliandsoon}. 

The physical picture behind the polarization dependence can be described as follows.
The interaction operator ${\mathbf A}\cdot{\mathbf p}$ has even (+) parity in a {\it parallel}
and odd (-) parity in a {\it perpendicular} experimental geometry. 
Assuming the applicability of the Zhang-Rice singlet state construction to the first
electron-removal final state, 
in a many-body picture this state belongs to the $A_{1g}$ representation, therefore being totally
 symmetric.
Thus in this representation the Zhang-Rice singlet has even parity with respect to a mirror plane along the Cu3$d_{x^2-y^2}$-O2$p_x$/O2$p_y$
orbital  bonds ($M_2$) as well as at $45$ degrees to the bonds ($M_1$).
In $\mathbf{k}$-space $M_1$ corresponds to the \Gpipi\ direction and $M_2$ to
\Gpino.
The initial state (ground state) is a one-hole state with $d_{x^2-y^2}$-symmetry,
and  therefore has even parity with respect to $M_2$ and odd parity with respect to $M_1$.
The matrix element thus formally vanishes for the two cases

\begin{eqnarray*}
\mbox{parallel for \Gpipi}&\Rightarrow &M\sim\langle +|+|-\rangle=0\\
\mbox{perpendicular for \Gpino}&\Rightarrow &M\sim\langle +|-|+\rangle=0
\end{eqnarray*}

This argumentation also holds in the one band picture.
Here the initial state (ground state) is totally symmetric and the final state has
$d_{x^2-y^2}$-symmetry (as shown in Fig. \ref{mirror}), leading to the same result.  

Thus, the observed polarization dependence of the first electron-removal states
of an undoped CuO$_2$-plane indicates that these states have a symmetry fully consistent with that 
of the Zhang-Rice singlet.
These results amend earlier reports (Ref.  \onlinecite{Hayn99})
regarding the polarization dependence of the first electron-removal states along \Gpino\ in
Sr$_2$CuO$_2$Cl$_2$.

\section{The dispersion relation and spectral function of the first electron-removal states}

The common picture given by the \tJm\ and its
extensions \cite{Trugman90,Liu92,Dagotto94,Moreo95} is a
strong dispersion of the Zhang-Rice singlet state along the \Gpipi\ direction 
with the minimum binding energy at \pizpiz.
At this $\mathbf{k}$-point, the spectral weight of the first electron-removal 
state also has its maximum and
vanishes going away from \pizpiz. Recent ARPES experiments on
Sr$_2$CuO$_2$Cl$_2$ confirm this \cite{Wells95,LaRosa97,Kim98,Haffner00,Pothuizen97}.
Along the \Gpino\ direction the \tJm\ predicts a rather low binding
energy near the high-symmetry point \pino\ which is, in fact, almost energetically 
degenerate with that at \pizpiz.
In the same model, the spectral weight of the quasiparticle
increases and is maximal at \pino. This is in contrast to the result
given in Ref. \onlinecite{Wells95} where no dispersion was observed and in
Refs. \onlinecite{LaRosa97,Kim98,Haffner00}
where the binding energy of the first electron-removal 
 state increases and the quasiparticle weight decreases after \mbox{($\pi/2$,$\,0$)}.
For the \pinonopi\ direction, the \tJm\ predicts only little dispersion, whereas experiment 
has shown a strong isotropic dispersion around the \pizpiz\ point
(Ref. \onlinecite{Wells95}).
Extensions of the \tJm\ and the spin and charge separation ansatz exhibit new
properties along \Gpino\ and \pinonopi. The Zhang-Rice singlet state around \pino\ can now
lower its energy by delocalization and is thus pushed to higher binding
energies and the dispersion in this direction is essentially parabolic.
In these models,
both a quasiparticle dispersion along \pinonopi\ which is isotropic
around \pizpiz\, as well as a reduced spectral weight near \pino\ can be achieved.
In the spin and charge separation model the isotropy along \Gpipi\ and \pinonopi\ is intrinsic
and near \pizpiz\ the dispersion relation is wedge-like rather than parabolic.
Surprisingly, the data in Ref. \onlinecite{Wells95} agree well with the extended \tJm\
along \pinonopi, but not along \Gpino. 
In contradiction to that, other data along \Gpino\ are consistent with the extended \tJm\
and the spin and charge ansatz  \cite{LaRosa97}, a view which appears to be supported by 
further data sets \cite{Kim98,Haffner00}.
Thus, the experimental dispersion along the \Gpino\ remains controversial, and
yet it is the dispersion and the evolution of the spectral weight along
\Gpino\ that is of the deepest theoretical interest for the following reasons.
Firstly, the states near \pino\ evolve with increasing hole concentration to become the
flat bands located near the chemical potential in the metallic systems.
These flat band regions are believed by some to hold the key to high temperature
superconductivity.
Secondly, the dispersion and quasiparticle spectral weight along \Gpino\ are fingerprints
for the different models and parameters used in them.
Thirdly, there are strong indications that the dispersion in the
insulator is closely connected with the pseudogap in the underdoped region of the
high $T_c$ superconductor phase diagram and thus may also be related to the superconducting
 gap in the doped
superconducting systems  \cite{Laughlin97,Ronning98}.

Consequently, the importance of experimental data which allow the determination of the
dispersion and spectral weight of the first electron-removal  states along \Gpino\ cannot be
 overestimated. 
Nevertheless, the experimentalist is faced with a number of challenges when collecting
data along this direction:
firstly, at several points in the Brillouin zone, there is no or only weak
photointensity from the first electron-removal  state.
For example, near $\Gamma$ the Zhang-Rice singlet has no spectral weight,
because the Cu3$d_{x^2-y^2}$ and 
O2$p_x$/O2$p_y$ orbitals cannot hybridize there  \cite{Pothuizen97}, in addition the 
matrix elements for emission along the surface normal are formally zero for a perfectly
two-dimensional electronic state located in the CuO$_2$-plane.
Furthermore, near \pino\ the photoemission intensity of the first electron-removal states
becomes weak as predicted by the extended \tJm s and the spin and charge separation
ansatz.
Secondly, only perfectly aligned samples with fresh and clean surfaces excited with
\mbox{UV-light} in a well-adjusted measurement geometry will show usable photoemission 
data from the first electron-removal  state.
Lastly, as the clean, defect-free surface of Sr$_2$CuO$_2$Cl$_2$ is highly insulating,
avoiding charging effects can be difficult.
 
In Fig. \ref{disp-series} we present series of ARPES spectra recorded along \Gpino\ taken
with parallel polarization and \Gpipi\ with perpendicular polarization and with
photon-energies of $16$, $22$ and $35$ eV - thus measuring near a peak in the photoemission 
intensity oscillation in each case. 
The position of the low energy part of the dispersive structure has been fitted using a Gaussian 
function while the high binding-energy background was modelled simply with the tail of another 
Gaussian. Although this fit is physically somewhat simplistic, it proved to be sufficient to
reliably find the positions of the peak as indicated by the triangles in each plot. 
The absolute value on a binding-energy scale changes from one cleavage to
the next and is rather arbitrary due to the insulating nature of the substance.  
However, the two series with $16$ eV were taken during one and the same cleavage, which is the 
prerequisite to compare the binding-energy scales of both series.
The binding-energy scales of the $22$ and $35$ eV data have been adjusted
to that of the $16$ eV data.
We present a summary of the dispersion relations derived from the fits to the data in Fig.
\ref{dispersion}. 
\par
We repeat here that we observe no evidence for a photon-energy dependence of the dispersion 
relation of the first ionization states, as is consistent with the 
arguments regarding the spectral function given in the introduction. 
The dispersion of the first electron-removal states along \Gpipi\ agrees with the previous 
results having a parabolic shape with its minimum binding energy at \pizpiz. 
Along \Gpino\ we also find a parabolic dispersion, in agreement with
the dispersion reported in Refs. \onlinecite{LaRosa97,Kim98,Haffner00},
but in contrast to the data given in Ref. \onlinecite{Wells95}.
For $16$ eV photon-energy, we measured 72 meV difference in the lowest binding energies
of the first electron-removal state along each of the two high-symmetry directions,
which is half as small as in Ref. \onlinecite{LaRosa97}.
In Ref. \onlinecite{Kim98}, a series of spectra representing a ``maximum intensity cut'' for a photon-energy $22.4$ eV along a line from \pizpiz\ to \mbox{($0.67\pi$,$\,0$)} was shown,
which displayed a dispersion of about $300$ meV.
However, as we will show, the points of maximum intensity 
along \Gpino\ do not coincide with those possessing minimum binding energy.
Thus such a maximum cut will not trace the line of minimum binding energy.
It turns out that this difference strongly influences the parameters in model
Hamiltonians fitted to the dispersion curves.

\par
We now go on to compare the experimental dispersion with theoretical results
obtained within the extended \tJm
\begin{equation}\label{extendedtJm}
H=-\sum_{i\sigma}\,\sum_{l=1}^3\,t_l\,X_i^{\sigma 0}\,X_{i+l}^{0 \sigma}
+J\,\sum_{\langle i,m\rangle}\,\mathbf{S}_i\cdot\mathbf{S}_m\quad ,
\end{equation}
with the additional hopping terms to second ($t_2$) and third ($t_3$) neighbors
now added besides the dominating nearest neighbor hopping $t=t_1$. The Hamiltonian is
written in terms of Hubbard operators  $X_i^{\sigma
0}=c_{i\sigma}^\dagger(1-n_{i-\sigma})$ where $\sigma=\pm 1$ is the spin
index.
The values of the hopping terms are determined by mapping the more realistic
Emery model with parameters derived from a constrained density-functional
calculation  \cite{Hybertsen90} to its low binding-energy part  by means of the
cell-perturbation method. 
This procedure gives $t_2/t_1=-0.08$ and $t_3/t_1=-0.15$.
The theoretical dispersion relation shown in Fig. \ref{dispersion} is calculated using a
variational method involving a spin polaron of small radius \cite{Hayn97}.
There are also related works which lead to similar dispersions\cite{Nazarenkoandso}.

\par
It has been shown in exact diagonalisation studies of small clusters \cite{Dagotto90} that
the dispersive bandwidth scales with the single parameter, $J$, whereas the form of the 
dispersion curve is fixed by the ratios $t_2/t_1$ and $t_3/t_1$.
Therefore we use the variational ansatz for $J=t_1$ and then we scale the
bandwidth in Fig. \ref{dispersion} with a factor \mbox{$J=0.22$ eV}.
For reference, we also show the dispersion relation without
additional hopping terms ($t_2=t_3=0$, scaling factor \mbox{$J=0.28$ eV}),
which shows a much too small energy difference between the lowest binding
energies of the first electron-removal states along \Gpino\ and \Gpipi\ to obtain good
agreement with the experimental spectra. 

\par
Another important prediction from model Hamiltonians is
the evolution of the spectral weight along certain directions in the Brillouin zone.
Here we map this evolution and distinguish between the coherent and 
incoherent parts of the first electron-removal states following the procedure
proposed in Ref.  \onlinecite{Leung97}, in which the contribution from the main valence 
band tail was first subtracted from the spectra.
The photoemission intensity was then defined as the coherent 
spectral weight and fitted using a Gaussian.
The ARPES intensity which is neither part of the valence band tail
nor in the Gaussian is then taken to be a measure of the incoherent part of
the first electron-removal state spectral function [see Ref. \onlinecite{Leung97}].
In our case, we have used a Lorentzian broadened by the experimental resolution
({\it i.e.} a Voigt profile) to fit the lowest binding energy intensity (the coherent part).
Of course, this procedure is not physically rigorous, but does offer a rough estimation 
of the possible split between the coherent and incoherent spectral weight.

\par
In Fig. \ref{sweight}, the momentum distributions of the two parts of the ARPES
spectra are shown for $16$ eV and $22$ eV excitation energy.
Note that the overall shape of these distributions is independent of the photon-energy.
Along \Gpipi\ both the coherent and incoherent parts of the first electron-removal
states are symmetric around \pizpiz\, at which point they both reach their maximum weight.  
In the \Gpino\ direction, we observe a steady increase in spectral weight of both components 
up to \mbox{($0.7\pi$,$\,0$)} after which point it drops fast.
One can find signs of a qualitatively similar behavior in ARPES data from
other groups \cite{Wells95,LaRosa97,Kim98,Haffner00,Haffner00b}
and for measurements on the related substance
Ca$_2$CuO$_2$Cl$_2$ (Ref. \onlinecite{Ronning98}), although the coherent and incoherent
spectral weights were not analysed in these studies. 
Note that along \Gpino\ the minimum binding energy of the low binding-energy feature and the
maximum of the coherent intensity do not coincide at the same $\mathbf{k}$-point.
In calculations using extensions of the \tJm\ this behavior was predicted
\cite{Eskes96,Kyung96} and can now be considered as experimentally verified.

\par
Within the simple fit procedure we can give these quantities some numbers: the ratio of the coherent to incoherent 
spectral weight for the two locations in $\mathbf{k}$-space is 0.6 (\pizpiz) and 0.5 (\mbox{($0.7\pi$,$\,0$)}) for 16 eV photon-energy 
and 2.4 (\pizpiz) and 3.8 (\mbox{($0.7\pi$,$\,0$)}) for 22 eV photon-energy.
It is important to realise that we observe here an extremely strong apparent dependence of the {\it ratio} between the coherent and incoherent spectral weight upon the photon-energy.
To discuss the physical significance of this, we re-write equation \ref{spectralweight} as
\begin{equation}\label{coherentincoherent}
S({\mathbf k},E)=-\frac1\pi\,\mbox{Im}\,\langle i\mid
c_{\mathbf k}\,(\frac1{E-E({\mathbf k})-\Sigma}
+G_{\mbox{inc}})\,c^\dagger_{\mathbf k}
\mid i\rangle,
\end{equation}
in order to separate the coherent from the incoherent spectral weight
($\Sigma$ is the self-energy of the quasiparticle).
Plugging this into equation \ref{photointensity}, it is immediately clear that in taking the {\it ratio} of the coherent and the
incoherent intensity, the matrix element cancels out.
Thus, this ratio is then determined by the spectral function $S({\mathbf k},E)$ alone and should not depend on the excitation energy. 
\par
Our observation that the ratio does appear to depend on h$\nu$ leads then to the following possible conclusions:
(i) the fit procedure used is too simple and therefore does not correctly quantify, however roughly, the coherent and incoherent parts of the
spectral function.
(ii) there are significant extrinsic contributions to the photoemission signal in this binding energy region. These could be the result of an energetic
shift of spectral weight due to inelastic losses \cite{Joynt99}, which could be sensitively dependent on the photoelectron kinetic energy, as are the data from 
electron energy loss spectroscopy in reflection of Sr$_2$CuO$_2$Cl$_2$ (Ref. \onlinecite{PothuizenPhD}).
(iii) the photoemission intensity between the the main valence band edge and the chemical potential is not derived from 
the spectral weight (coherent + incoherent) resulting from a {\it single} electronic state - i.e. there is spectral weight from an additional, different
electronic state in this energy region. 
\par
The fit procedure, as we have discussed above, {\it is} simplistic - however the coherent-incoherent intensity ratios vary by a factor of more than four
between the two photon-energies.
The third possible conclusion - that there has to be spectral weight from more than one state in this energy region could have important implications.
We can rule out a significant contribution from secondary electrons for the relatively high kinetic energies dealt with here.
Intensity from surface states is unlikely, as XPD data \cite{Boeske97} shows that the cleavage surface of Sr$_2$CuO$_2$Cl$_2$ is
terminated within the SrCl units, which are essentially ionic and therefore do not support electronic states close to the chemical potential.
In addition, our LEED investigations also gave no evidence for a reconstruction of this termination layer.
Spin-resolved resonant photoemission of CuO (Ref. \onlinecite{Tjeng97})
has shown intensity due to triplet states within 1eV of the Zhang-Rice singlet, although
those states would be expected to have the same photon-energy dependence as the singlet in our experiment.

\par
If there was an additional, different state in this energy region, this would lead to a complicated $\mathbf{k}$- and h$\nu$-dependent overlap between
the intensity of this state and the higher binding energy components attributed up till now to incoherent weight from the Zhang-Rice singlet.
While this would not be expected to have a large impact on the observed dispersion relation for the Zhang-Rice singlet (which is, after all, a quantity derived from 
the spectral structure at lowest binding energies for each $\mathbf{k}$-point), it could, however lead to the observed photon-energy dependence of the ratio
between the low binding-energy (``coherent'') and higher binding-energy (``incoherent'') parts of the photoemission spectra.

\section{Summary}

In conclusion, we have presented a detailed ARPES study of the low binding-energy
occupied electronic structure of Sr$_2$CuO$_2$Cl$_2$, which corresponds to an investigation of the first electron-removal states of an undoped CuO$_2$-plane.
Our experiments, and the comparison of their results with theoretical models, have
revealed the following main points:

1. The photoemission signal of the first electron-removal states at both \pizpiz\
 and \mbox{($0.7\pi$,$\,0$)} exhibits a marked photon-energy dependence.
The intensity profile shows strong oscillations with maxima near $16$, $25$, $35$ and 
$49$ eV, corresponding to final state crystal momenta
k$_\perp$=$0.82$, $1.63$, $2.40$ and $3.12$ \AA$^{-1}$.
This strong photon-energy dependence has complicated comparisons between data in the
literature from different groups as regards both the spectral weight
and spectral form of the first electron-removal  states in these systems.
We attribute the oscillation of photoemission intensity (which has a period in k$_\perp$ of ca. 
$0.75$ \AA$^{-1}$) to the diffraction of the photoelectron 
wave on the periodic c-axis separation of the CuO$_2$-planes of $8.2$ \AA.

2. Along the high-symmetry directions \Gpipi\ and \Gpino\ the first electron-removal states
shows a strong polarization dependence. This can be linked to the strongly
polarization-dependent matrix element, which in turn allows the determination of 
the symmetry of the first electron-removal state itself.
For both high-symmetry directions we observe a polarization dependence in keeping with 
that expected for a Zhang-Rice singlet state in the framework of either a three-band or one-band model Hamiltonian. 

3. Our data show that the dispersion of the first electron-removal states along both high
symmetry directions (\Gpipi\ and \Gpino) is parabolic-like and independent
of the excitation energy.
This, and the rather large difference in lowest binding energy of the first electron-removal  
state along these directions, shows the validity of the extended \tJm\ for describing the disperion 
relation of a single hole in an antiferromagnetic CuO$_2$ plane. 
Thus, the inclusion of second ($t_2$) and third ($t_3$) neighbor hopping terms with realistic 
values of $t_2=-0.08$ and $t_3=0.15$ in units of the next neighbor hopping $t=t_1$ are required.

4. Upon application of a simple fit procedure, we infer the momentum distribution of the spectral weight of the coherent and incoherent part of the first electron-removal  state to have its maximum 
along \Gpipi\ at \pizpiz\, being symmetrically suppressed away from this point.
Along \Gpino\ the spectral weights of both parts reach their maximum at
\mbox{($0.7\pi$,$\,0$)} and then drop fast.
The ratio between the coherent and incoherent spectral weight is strongly photon-energy dependent, which, at
first sight would appear to violate the physics of the spectral function.
There are different possible explanations for this including: (i) the necessity for a more 
sophisticated framework in which to analyse the weight of the coherent and incoherent contributions to the spectral weight;
(ii) significant (h$\nu$-dependent) intensity due to extrinsic processes
(iii) intensity in this energy region due to intrinsic electronic states {\it other} than the Zhang-Rice singlet.

\par
We gratefully acknowledge the stimulating conversations with S. Haffner.
This work was supported by the BMB+F under contract number 05 SB8 BDA6 and
by the DFG under Fi439/7-1 and as part of the Graduiertenkolleg ``Struktur und Korrelationseffekte im Festk\"orper'' der TU Dresden.

\begin{figure}
\caption{\label{mirror}
Schematic representation of the Zhang-Rice singlet state in the one-particle representation. The Zhang-Rice singlet
has the same symmetry as a $d_{x^2-y^2}$-orbital, implying the existence of  two mirror planes perpendicular to the CuO$_2$-plane, labelled $M_1$ and $M_2$.
}
\end{figure}

\begin{figure}
\caption{\label{geometry}
The experimental geometry for the photoemission experiments: a) emission-plane and polarization-plane are parallel to each other, b) emission-plane and polarization-plane are perpendicular to each other.
Throughout this paper we call the first case a) {\it parallel} and the second case b) {\it
perpendicular} polarization.}
\end{figure}

\begin{figure} 
\caption{\label{hnscan1}
A typical example of ARPES data of the 
first electron-removal states of Sr$_2$CuO$_2$Cl$_2$ as a function of photon-energy.
In panel a) the ARPES EDCs are recorded at \pizpiz\ and in panel b) at
\mbox{($0.7\pi$,$\,0$)}. The photon-energies are as shown.}
\end{figure}

\begin{figure}
\caption{\label{hnscan2}
The photon-energy dependence of the photoemission intensity of the
first electron-removal states a) at \pizpiz\ and b) at \mbox{($0.7\pi$,$\,0$)},
arrived at from the analysis of a large body of ARPES data such as that shown in
Fig. \ref{hnscan1}.
The solid black line represents a guide to the eye and the error bars
(located in the upper left corner of each panel) indicate the error of the
fit to the experimental intensity (95\% confidence interval).
The top axis shows the corresponding k$_\perp$ scale calculated using
an inner potential of $8.0$ eV (for details see text).}
\end{figure}

\begin{figure}
\caption{\label{series}
Series of energy-distribution curves for different polarization geometries. The
photon-energy was set to 22.4 eV and all  spectra were taken at 300K. Panels a) and
b) show measurements along \Gpipi\ where the first electron-removal state peak appears near
$21.0$ eV kinetic energy  for perpendicular polarization in a). The
corresponding series along \Gpino\ are shown in panels c) and d). There, the
first electron-removal states gives significant photocurrent near $20.9$ eV kinetic energy only for parallel polarization in d).
}
\end{figure}

\begin{figure}
\caption{\label{disp-series}
ARPES series showing the dispersion of the Zhang-Rice singlet state along the high-symmetry
directions. The upper three panels show series along \Gpipi\, the lower three panels along \Gpino . The excitation energy is the same for each column, $16$ eV (left), $22$ eV (center) and $35$ eV (right). Black triangles are included as a guide indicating the kinetic energy of the
first ionization state determined by a fitting procedure described in the text.
}
\end{figure}

\begin{figure}
\caption{\label{dispersion}
The dispersion of a single hole in an antiferromagnetic CuO$_2$-plane
measured along two directions in $\mathbf{k}$-space. 
The symbols represent data from the ARPES spectra from Sr$_2$CuO$_2$Cl$_2$, recorded with 
the indicated photon-energies.
The vertical error bars in the right panel indicate the errors of the fit (95\% confidence interval) in determining the binding energy of the first ionization states.
The $\mathbf{k}$-resolution is $0.054$, $0.075$ and $0.094$ \AA$^{-1}$ for 16, 22 and 35 eV
photon-energy respectively, which corresponds
to 4.5\%[6.8\%], 6.3\%[9.6\%] and 7.8\%[11.9\%] of the direction \Gpipi [\Gpino ].
The lines indicate the results of the calculations employing a
variational ansatz using a spin polaron with small radius (for details see text).
The dashed line corresponds to a calculation where the additional transfer terms $t_2$ and
$t_3$ are chosen to be zero (i.e. the \tJm ), while for the solid line these terms are set
to $t_2/t_1=-0.1$ and $t_3/t_1=0.2$ (i.e. an extended \tJm). 
The improved agreement to the experimental data provided by the extended \tJm,
particularly along \Gpino\ is apparent.}
\end{figure}

\begin{figure}
\caption{\label{sweight}
The coherent and incoherent part of the spectral weight of the first electron-removal
states determined from the data shown in Fig. \ref{disp-series} for a) 22 eV and b)
16 eV photon-energy following the procedure used by Leung {\it et al.},
Phys. Rev. B {\bf 56}, 6320 (1997).
The upper part in each case shows a representative result of the fit. 
In the lower part of the figure,
the weight of the coherent (filled circles) and the incoherent (open circles) intensity,
normalized to the coherent spectral weight at \pizpiz, is shown. The vertical bars indicate
the error of the least squares fit to the ARPES intensity (95\% confidence interval).
}
\end{figure}


\begin{references}
\bibitem{fer-zrs}Throughout this paper we strictly differentiate between the
term first electron-removal  state, which is a real object, and the Zhang-Rice singlet state,
which is rather a
theoretical construct to describe the first electron-removal  state.
\bibitem{brillouin}We refer to the notation in Ref. \onlinecite{Wells95}
where $\Gamma=(0,0)$ means the center and $(\pm\pi,\pm\pi)$ means the
corners of the two-dimensional Brillouin zone of the square-planar CuO$_2$-plane
given in units $1/a$ where $a=3.967$ \AA is the lattice constant of
Sr$_2$CuO$_2$Cl$_2$ (Ref. \onlinecite{Miller90}).
\end{references}
\end{document}